\documentclass[fp,twocolumn]{jpsj3}%
\usepackage{amsmath}
\usepackage{txfonts}
\usepackage{amsfonts}
\usepackage{amssymb}
\usepackage{graphicx}
\usepackage{bm}
\usepackage{color}%
\providecommand{\U}[1]{\protect\rule{.1in}{.1in}}
\inst{$^1$Division of Natural and Environmental Sciences, The Open University of
Japan, Chiba 261-8586, Japan \\
$^{2}$ Department of Physics and Astronomy, University of Manitoba, Winnipeg MB R3T 2N2, Canada \\
$^{3}$ Institute of Natural Sciences and Mathematics, Ural Federal University, Ekaterinburg
620002, Russia\\
$^{4}$ Institute for Molecular Science,
38 Nishigo-Naka, Myodaiji, Okazaki, 444-8585, Japan}
\abst{An optical vortex beam carries orbital angular momentum $\ell$ in addition to spin angular momentum $\sigma$. We demonstrate that  a Landau-quantized two dimensional electron system absorbs the optical vortex beam through modified selection rules, reflecting two kinds of angular momenta.  The lowest Landau level electron absorbs the optical vortex beams with $\sigma=1$ (positive helicity) and $\ell=0$ or $\sigma=-1$ (negative helicity) and $\ell=2$ in the electric dipole transition. The induced electric currents survive only along the edge of the sample, due to cancellation of the bulk currents. Thus, the magnetization can be induced by only the edge current. It is shown that the induced orbital magnetization disappears when the dark ring of the beam coincides with the disk edge. This scheme may provide a helicity-dependent absorption using the optical vortex beam. }
\begin{document}

\title{Selection Rules for Optical Vortex Absorption by Landau-quantized Electrons}
\author{Hirohisa T. Takahashi,$^{1}$\thanks{1517000122@campus.ouj.ac.jp}, Igor
Proskurin,$^{2,3}$, and Jun-ichiro Kishine$^{1,4}$}
\maketitle

\section{Introduction}

Originally, it was suggested by Poynting that circularly polarized light
carries spin angular momentum (SAM) equal to $\pm\hbar$ per photon, which can
be transferred to medium and produce a mechanical torque in light-matter
interactions. \cite{Poynting1909} Later, Beth experimentally confirmed angular
momentum transfer from light in 1935. \cite{Beth1935,Beth1936} After about a
century, it was suggested that lights can also carry an orbital angular
momentum (OAM) in addition to SAM by Allen \textit{et al..}\cite{Allen1992}
This part of angular momentum appears as a modulation of a phase front, so it
was dubbed an optical vortex (OV) or twisted light. It was experimentally
demonstrated that a single photon is able to carry quantized OAM.
\cite{Mair2001} Theoretical and experimental techniques were developed to
generate OVs in various forms such as the Laguerre-Gaussian (LG) and the
Bessel light beams. \cite{Allen1992,Durnin1987a,Andrews2008Book} These unique
forms of light beams have triggered much interest on the transfer of optical
OAM to material particles and atoms via light-matter interactions.
\cite{Sonja2016,Padgett2017}

Mathematically, OV is described by a constant phase profile given by
$\exp(i\ell\varphi+ikz)$, where $\varphi$ is the azimuthal angle in the
cylindrical coordinate system for a light beam propagating in the $z$
direction with the wavenumber $k$. It carries an intrinsic OAM equal to
$\pm\ell\hbar$ per photon ($\ell=0,\pm1,\pm2,\ldots$), which is independent of
the polarization state of light. \cite{Allen1992} Geometrically the phase
front of OV is a helix with the winding number determined by $\ell$. The
radial dependence of the beam amplitude is typically given in terms of either
Laguerre-Gaussian or Bessel modes. The former has the property of gradually
expanding as the beam propagates, while the latter is diffraction free, or
propagation invariant.\cite{Durnin1987a,Durnin1987b} Experimentally, Bessel
OVs can be created in the back focal plane of a convergent lens by a plane
wave,\cite{Durnin1987b} by an axicon lens from a Gaussian
beam,\cite{Indebetouw1989} by the use of computer-generated holograms,
\cite{Vasara1989} or by a Fabry-Perot resonator. \cite{Cox1992}

From the point of classical mechanics, exerting a torque by transferring
angular momentum from OV has been actively studied, for example, with
particles rotating in an optical
tweezers,\cite{Amos1995,Friese1995,Friese1998} and the laser ablation
technique. \cite{Hamazaki2010} In recent years, coupling of twisted light with
condensed matter also saw a considerable development, including such topics as
generation of atomic vortex states by coherent transfer of OAM from photons to
the Bose-Einstein condensate \cite{Andersen2006}, photocurrents excited by the
OV beam-absorption in semiconductors and
graphene,\cite{Quinteiro2009a,Watzel2012,Quinteiro2013} excitation of
multipole plasmons in metal nanodisks,\cite{Sakai2015} spin and charge
transport on a surface of topological insulator,\cite{Shintani2016} generation
of skyrmionic defects in chiral magnets.\cite{FujitaSato2017}

However, whether OAM affects any spectroscopic selection rules via optically
induced electronic transitions is still an open question. Although
transferring of the OAM to atomic electrons from the OV beam via the electric
quadrupole transition was
reported,\cite{Schmiegelow2016,Afanasev2018,Solyanik2019} for electric dipole
transitions in atoms, it has been proved that the optical OAM can be
transferred only to the center-of-mass motion of the atom or
molecule,\cite{Babiker2002,Lloyd2012} thus the electric dipole selection rules
remain unchanged. Similar in the coupling of OV with the exciton, the optical
OAM can be transferred only to the exciton center-of-mass
motion.\cite{Shigematsu2016} These phenomena are analogous to the fact that
the cyclotron resonance frequency is independent of short-range
electron-electron interactions.\cite{Kohn1961}

In this case, it interesting to see whether these concepts are applicable to a
degenerated two-dimensional electron gas (2DEG) in magnetic field. To our best
knowledge, such a system has not been considered before our previous
letter,\cite{Takahashi2018} where we discussed the optical conductivity and
the selection rules in 2DEG exposed to OV with optical OAM. By applying the
magnetic field, 2DEG is characterized by discrete energy levels with localized
semi-classical electron orbits. It was demonstrated that the bulk current
induced by OV disappears, and only the edge current survives when the 2DEG is
irradiated by a Bessel beam.\cite{Takahashi2018} This situation is similar to
the picture of orbital magnetization, \cite{Thonhauser2005} which is known to
appear due to the existence of the edge currents. Therefore, in 2DEG we can
anticipate an orbital Edelstein effect \cite{Yoda2018} where additional
magnetization is induced by the OV, which is the central issue of this paper.
In this paper, we extend discussions on the results we shortly presented in
our previous letter to present theoretical details including the induced
orbital magnetization.\cite{Takahashi2018}

This paper is organized as follows. We briefly review the derivation of a
circularly-polarized Bessel-mode OV in Section II and 2DEG on circular disc in
Sec. III. We calculate the induced photocurrent in 2DEG in Sec. IV using the
Kubo linear response theory. In Sec. V, we discuss cancellation of bulk
currents in the semi-classical picture. It is demonstrated how the
magnetization is induced by the OV light beam in Sec. VI. Sec. VII is reserved
for conclusions.

\section{Circularly Polarized Bessel-mode Optical Vortex}

It is crucial for studying quantum mechanical properties of light to separate
the total angular momentum (TAM) into spin and orbital parts, since they can
be conserved separately for light interacting with particles. In the paraxial
approximation, this separation can be done explicitly, and the light beam has
a well-defined SAM related to its polarization state and OAM determined by the
phase modulation. In this paper, we adopt the paraxial approximation, which is
a usual situation in real-world experiments.

We briefly review derivation of the Bessel-mode OV within the paraxial
approximation following Refs. \citen{Matura2013,Jentschura2011} The wave
equation for the vector potential of a monochromatic light $\mathbf{A}\left(
\mathbf{r},t\right)  =\mathbf{A}\left(  \mathbf{r}\right)  \mathrm{e}%
^{-i\omega t}$ with the frequency $\omega$ in vacuum in the Coulomb gauge is
given by the Helmholtz equation:
\begin{equation}
\Delta\mathbf{A}\left(  \mathbf{r}\right)  +k^{2}\mathbf{A}\left(
\mathbf{r}\right)  =0, \label{Helmholtz eq}%
\end{equation}
where $\Delta$ is a Laplace operator, and $k^{2}=\omega^{2}/c^{2}$ with a
speed of light in a vacuum $c$. In order to obtain twisted solutions, we have
to take account of additional requirements. First is that $\mathbf{A}\left(
\mathbf{r}\right)  $ is a propagating wave along $z$-axis, so it is the
eigenvector of the linear momentum operator $p_{z}=-i\hbar\nabla_{z}$,
$\hat{p}_{z}\mathbf{A}\left(  \mathbf{r}\right)  =\hbar k_{z}\mathbf{A}\left(
\mathbf{r}\right)  $. Second is that $\mathbf{A}\left(  \mathbf{r}\right)  $
should also be the eigenvector of $z$-component of the TAM operator
\begin{equation}
\hat{J}_{z}\mathbf{A}\left(  \mathbf{r}\right)  =J\mathbf{A}\left(
\mathbf{r}\right)  ,
\end{equation}
where the operator $\hat{J}_{z}=\hat{L}_{z}+\hat{S}_{z}$ is given by the
corresponding components of the orbital and spin angular momentum operators:%
\begin{equation}
\hat{L}_{z}=-i\hbar\frac{\partial}{\partial\varphi},\ \ \ \hat{S}_{z}=-i\hbar%
\begin{pmatrix}
0 & 1 & 0\\
-1 & 0 & 0\\
0 & 0 & 0
\end{pmatrix}
, \label{OAMSAMop}%
\end{equation}
where we define the modulus of the transverse linear momentum $k_{\perp
}=\left\vert \mathbf{k}_{\perp}\right\vert =\sqrt{k^{2}-k_{z}^{2}}$.

The normalized scalar solution of the Helmholtz equation in cylindrical
coordinates $\left(  r_{\perp},\varphi,z\right)  $ can be written in the form
\begin{equation}
\psi_{\ell}\left(  \mathbf{r}|k_{\perp},k_{z}\right)  =\sqrt{\frac{k_{\perp}%
}{2\pi}}J_{\ell}\left(  k_{\perp}r_{\perp}\right)  \mathrm{e}^{i\ell\varphi
}\mathrm{e}^{ik_{z}z},
\end{equation}
where $\ell$ determines the OAM of light which is the eigenvalue of the OAM
operator (\ref{OAMSAMop}), and $J_{n}(x)$ is the $n$-th order Bessel function
of the first kind. The normalization condition is
\begin{align}%
%TCIMACRO{\dint }%
%BeginExpansion
{\displaystyle\int}
%EndExpansion
\psi_{\ell^{\prime}}^{\ast}\left(  \mathbf{r}^{\prime}|k_{\perp}^{\prime
},k_{z}^{\prime}\right)  \psi_{\ell}\left(  \mathbf{r}|k_{\perp},k_{z}\right)
d^{3}r  &  =2\pi\delta\left(  k_{\perp}-k_{\perp}^{\prime}\right) \nonumber\\
&  \times\delta\left(  k_{z}-k_{z}^{\prime}\right)  \delta_{\ell,\ell^{\prime
}}.
\end{align}

Expansion over plane waves of the scalar function $\psi_{\ell}\left(
\mathbf{r}|k_{\perp},k_{z}\right)  $\ is%
\begin{align}
\label{eq6}\psi_{\ell}\left(  \mathbf{r}|k_{\perp},k_{z}\right)   &
=\int\frac{d^{2}\mathbf{k}_{\perp}^{\prime\prime}}{\left(  2\pi\right)  ^{2}%
}a_{k_{\perp},\ell}(\mathbf{k}_{\perp}^{\prime\prime})\mathrm{e}%
^{i\mathbf{k}^{\prime\prime}\cdot\mathbf{r}}\nonumber\\
&  =\int\frac{d^{2}\mathbf{k}_{\perp}^{\prime\prime}}{\left(  2\pi\right)
^{2}}a_{k_{\perp},\ell}(\mathbf{k}_{\perp}^{\prime\prime})\mathrm{e}%
^{i\mathbf{k}_{\perp}^{\prime\prime}\cdot\mathbf{r}_{\mathbf{\perp}}%
+ik_{z}^{\prime\prime}z}%
\end{align}
with $\mathbf{k}_{\perp}^{\prime\prime}=(k_{\perp}^{\prime\prime}\cos
\varphi_{k},k_{\perp}^{\prime\prime}\sin\varphi_{k},0)$ and $\mathbf{r}%
_{\perp}=(\cos\varphi,\sin\varphi,0)$. Each plane wave component is written
as
\begin{equation}
a_{k_{\perp},\ell}(\mathbf{k}_{\perp}^{\prime\prime})=\sqrt{\frac{2\pi
}{k_{\perp}}}\left(  -i\right)  ^{\ell}\mathrm{e}^{i\ell\varphi_{k}}%
\delta\left(  k_{\perp}^{\prime\prime}-k_{\perp}\right)  .
\end{equation}
These expressions show that $\psi_{\ell}\left(  \mathbf{r}|k_{\perp}%
,k_{z}\right)  $ can be viewed as a superposition of plane waves with fixed
$k=\left\vert \mathbf{k}\right\vert =\sqrt{k_{\perp}^{^{\prime\prime}2}%
+k_{z}^{2}}$ whose direction belongs to the cone with the cone angle
$\theta_{k}=\tan^{-1}k_{\perp}^{\prime\prime}/k_{z}$.

When the scalar solution of the Helmholtz equation is considered as a
superposition of plane waves, it is important to study the polarization
structure of the plane wave with the propagation vector $\mathbf{k}$. The
vector potential of the plane wave has to be an eigenvector of the SAM
operator, $\hat{S}_{z}\mathbf{A}^{\text{pl}}\left(  \mathbf{r}\right)  =\hbar
S\mathbf{A}^{\text{pl}}\left(  \mathbf{r}\right)  $. For the plane wave
traveling along $\mathbf{k}=(0,0,k_{z})$, the spin angular momentum operators
$\hat{S}_{z}$ has the following eigenvectors:%
\begin{align}
\mathbf{\eta}_{0}  &  =%
\begin{pmatrix}
0\\
0\\
1
\end{pmatrix}
\ \ \text{for }S=0,\ \nonumber\\
\mathbf{\eta}_{\pm}  &  =\mp\frac{1}{\sqrt{2}}%
\begin{pmatrix}
1\\
\pm i\\
0
\end{pmatrix}
\ \ \text{for }S=\pm1,
\end{align}
and the vector potential is given by $\mathbf{A}^{\text{pl}}(\mathbf{r}%
)=\mathbf{\eta}_{\sigma}A_{0}\mathrm{e}^{ik_{z}z}$, where $A_{0}$ is a constant.

When the plane wave travels in arbitrary direction $\mathbf{k}$, which does
not necessary coincide with the $z$-axis, $\mathbf{k}=k(\cos\varphi_{k}%
\sin\theta_{k},\sin\varphi_{k}\sin\theta_{k},\cos\theta_{k})$, its
polarization vector $\mathbf{\varepsilon}_{k,\sigma}$ can be found from
original polarization vectors $\mathbf{\eta}_{\sigma}$ by rotating them with
rotation matrix
\begin{align}
\hat{R}_{k}  &  =\hat{R}_{\varphi_{k}}\hat{R}_{\theta_{k}}\nonumber\\
&  =%
\begin{pmatrix}
\cos\varphi_{k} & -\sin\varphi_{k} & 0\\
\sin\varphi_{k} & \cos\varphi_{k} & 0\\
0 & 0 & 1
\end{pmatrix}%
\begin{pmatrix}
\cos\theta_{k} & 0 & \sin\theta_{k}\\
0 & 1 & 0\\
-\sin\theta_{k} & 0 & \cos\theta_{k}%
\end{pmatrix}
,
\end{align}
which gives
\begin{equation}
\mathbf{\varepsilon}_{k,\sigma}=\hat{R}_{k}\mathbf{\eta}_{\sigma}%
=-\frac{\sigma}{\sqrt{2}}%
\begin{pmatrix}
\cos\varphi_{k}\cos\theta_{k}-i\sigma\sin\varphi_{k}\\
\sin\varphi_{k}\cos\theta_{k}+i\sigma\cos\varphi_{k}\\
-\sin\theta_{k}%
\end{pmatrix}
.
\end{equation}
Then the vector potential for the plane wave traveling along $\mathbf{k}$ is
given by%
\begin{equation}
\mathbf{A}^{\text{pl}}(\mathbf{r})=\mathbf{\varepsilon}_{k,\sigma}%
A_{0}e^{i\mathbf{k}\cdot\mathbf{r}},\ \mathbf{\varepsilon}_{k,\sigma}%
\cdot\mathbf{k}=0,
\end{equation}
where the Coulomb gauge is used and the polarization vector
$\mathbf{\varepsilon}_{k,\sigma}$ then describes photon carrying a helicity
$\sigma=\pm1$. We can expand $\mathbf{\varepsilon}_{k,\sigma}$ over the
orthonormal basis $\left\{  \mathbf{\eta}_{S}\right\}  _{S=0,\pm1}$ of the
eigenvectors of the SAM operator $\hat{S}_{z}$:%
\begin{equation}
\mathbf{\varepsilon}_{k,\sigma}=\sum_{S=0,\pm1}c_{S,\sigma}\mathrm{e}%
^{-iS\varphi_{k}}\mathbf{\eta}_{S},
\end{equation}
where the expansion coefficients are given by
\begin{equation}
c_{0,\sigma}=\frac{\sigma}{2}\sin\theta_{k},\ c_{\pm1,\sigma}=\frac{1}%
{2}\left(  1\pm\sigma\cos\theta_{k}\right)  .
\end{equation}
Now we can find the expression for the vector potential for OV based on the
expansion over the plane waves in Eq.~(\ref{eq6}) and taking into account that
each plane wave is characterized by its own polarization vector
$\mathbf{\varepsilon}_{k,\sigma}$:
\begin{align}
\mathbf{A}^{\text{OV}}\left(  \mathbf{r}\right)   &  =\mathbf{A}^{\text{OV}%
}\left(  \mathbf{r|}k_{\perp},k_{z},J,\sigma\right) \nonumber\\
&  =A_{0}\int\frac{d^{2}\mathbf{k}_{\perp}^{\prime\prime}}{\left(
2\pi\right)  ^{2}}a_{k_{\perp},J}(\mathbf{k}_{\perp}^{\prime\prime
})\mathbf{\varepsilon}_{k,\sigma}\mathrm{e}^{i\mathbf{k}_{\perp}^{\prime
\prime}\cdot\mathbf{r}_{\mathbf{\perp}}+ik_{z}^{\prime\prime}z},
\end{align}
where we introduced $J$ as the eigenvalue of the TAM operator $\hat{J}%
_{z}=\hat{L}_{z}+\hat{S}_{z}$. Integrating over $\mathbf{k}_{\perp}%
^{\prime\prime}$, we finally obtain the vector potential of the OV with Bessel
mode
\begin{align}
\mathbf{A}^{\text{OV}}\left(  \mathbf{r|}k_{\perp},k_{z},J,\boldsymbol{\sigma
}\right)   &  =A_{0}\sqrt{\frac{k_{\perp}}{2\pi}}\sum_{S=0,\pm1}\mathbf{\eta
}_{S}(-i)^{S}c_{S,\sigma}J_{J-S}(k_{\perp}r_{\perp})\nonumber\\
&  \times\mathrm{e}^{i\left(  J-S\right)  \varphi}\mathrm{e}^{ik_{z}z}.
\end{align}

In the paraxial approximation, we assume that the longitudinal momentum of the
photon is much greater than its transverse momentum, $k_{z}\gg k_{\perp}$, so
the expansion coefficients become $c_{S,\sigma}\approx\delta_{S,\sigma}$, and
we get the vector potential in the form:
\begin{align}
\mathbf{A}^{\text{OV}}(\boldsymbol{r}|k_{\perp},k_{z},\ell+\sigma,\sigma)  &
\sim\mathbf{\eta}_{\sigma}A_{0}\sqrt{\frac{k_{\perp}}{2\pi}}(-i)^{\sigma
}J_{\ell}(k_{\perp}r_{\perp})\nonumber\\
&  \times\mathrm{e}^{i\ell\varphi}\mathrm{e}^{ik_{z}z}%
\label{VectorPotentialInPA}\\
&  \equiv\boldsymbol{A}_{\ell,\sigma}^{\text{OV}}\left(  \boldsymbol{r}%
\right)  ,\nonumber
\end{align}
where we introduced a OAM quantum number, $\ell=J-\sigma$. Moreover, if we
take the limit $k_{\perp}\rightarrow0$ with $r_{\perp}$ being fixed, then the
Bessel function gives $J_{\ell}(k_{\perp}r_{\perp})\rightarrow\delta_{\ell,0}%
$, and we recover a plane wave solution with $J=\sigma$ propagating along the
$z$-axis.

\begin{figure}[h]
\centering\includegraphics[scale=0.65]{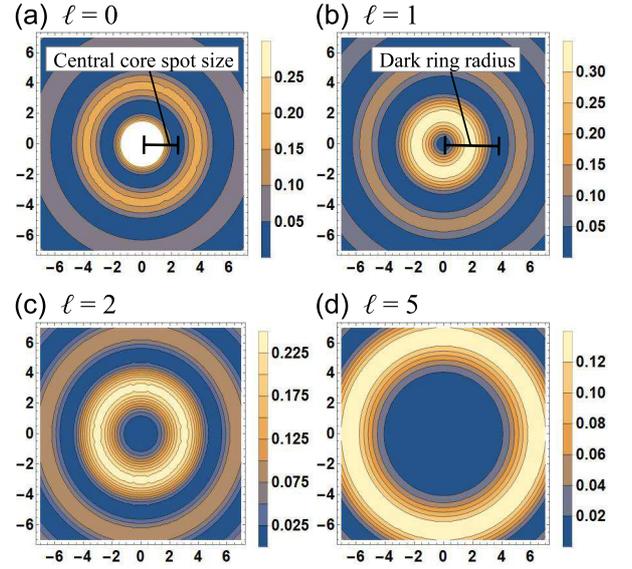}\caption{(Color online)
Some examples of the radial profile of the Bessel-mode optical vortex,
$\left\vert J_{\ell}(x)\right\vert ^{2}$. (a) $\ell=0$. The central core spot
size is given by the fisrt zeros of $J_{0}(x)$. (b) $\ell=1$, The dark ring
radius is given by $i$-th zeros of $J_{\ell}(x)$. (c) $\ell=2$, (d) $\ell=5$.}%
\label{Radial_profile}%
\end{figure}

The Bessel-mode OV exhibits a feature of being diffraction free and has a
phase singularity. The first feature can easily be seen by using Eq.
(\ref{VectorPotentialInPA}). The intensity of the vector potential,
$I\propto\left\vert \mathbf{A}\right\vert ^{2}$, is independent of $z$. The
phase singularity is located on the beam axis where the intensity becomes
zero. To demonstrate a transfer of OAM, the target particles are usually
located in non-zero intensity region off the beam axis and dark rings. The
radius of $i$-th dark ring of the higher-order Bessel beam is given by
\begin{equation}
r_{\perp}^{\ell,i}=\frac{(\text{the }i\text{-th zeros of }\ell\text{-th order
Bessel function})}{k_{\perp}}, \label{Dark_ring_radius}%
\end{equation}
which is determined by $J_{\ell}(k_{\perp}r_{\perp}^{\ell,i})=0$. In
particular, the central core size of the zero-order Bessel beam is given by
$r_{\perp}^{0,1}=2.404/k_{\perp}$. We exhibit some examples of the radial
profile of the Bessel-mode OV, and the definition of the dark ring radius and
the central core spot size in Figure \ref{Radial_profile}.We note that the Bessel-mode OV even with $\ell=0$ has the dark
rings corresponding to transversely traveling wave, This feature is also the
crucial difference with the plane wave.

\section{Landau-quantized Electron}

\begin{figure}[h]
\centering\includegraphics[scale=0.52]{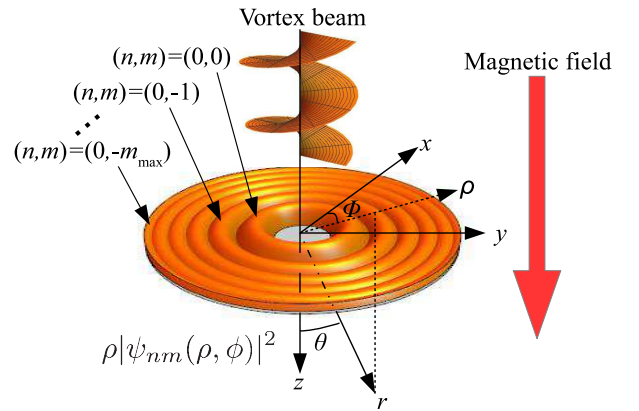}\caption{(Color online)
Schematic picture showing 2D electron distributions in the lowest LLs in the
circular disc geometry. The OV beam is vertically irradiated to 2DEG. The
direction of propagation of OV is taken the $z$-axis. Directions of the
induced photocurrents are indicated by arrows. The azimuthal angle $\varphi$
is on the 2D electron system. }%
\label{SetupConfig}%
\end{figure}

The quantized energy levels of 2DEG in the magnetic field $B$ are given by
\cite{LandauQM1977} $E_{N}=\hbar\omega_{c}\left(  N+1/2\right)  $, which
usually appear by solving the Schr\"{o}dinger equation in the Landau's gauge,
where $N=0,1,2,\dots$ is the Landau level index, and $\omega_{c}=eB/m_{e}$ is
the cyclotron frequency with the elementary charge $e\left(  >0\right)  $, and
the bare electron mass $m_{e}$. We here note that the electron mass $m_{e}%
$\ should be interpreted as an effective mass $m_{e}^{\ast}=0.067m_{e}$\ for
GaAs. However, when we consider 2DEG interacting with the Bessel OV light
beam, the symmetric gauge in the cylindrical coordinates becomes a natural
choice. Hence, we consider 2DEG on with a circularly shaped disk geometry and
take the cylindrical coordinates as shown in Fig.\ref{SetupConfig}.

The non-perturbative Hamiltonian for 2DEG under the magnetic field is given by
\begin{equation}
H_{0}=\frac{1}{2m_{e}}\left[  -i\hbar\mathbf{\nabla}+e\mathbf{A}^{\text{ext}%
}(\mathbf{r})\right]  ^{2} \label{Non-perturmativeHamil}%
\end{equation}
where $\mathbf{A}^{\text{ext}}(\mathbf{r})=\left(  -By/2,Bx/2,0\right)  $, and
the magnetic field is along the $z$-axis direction. The energy spectrum is
obtained by solving the Schr\"{o}dinger equation $H_{0}\Psi=E\Psi$ which
gives\cite{Darwin1931}%
\begin{align}
E_{n,m}  &  =\hbar\omega_{c}\left(  n+\frac{|m|+m}{2}+\frac{1}{2}\right)
,\nonumber\\
n  &  =0,1,2,...,\text{and }m=0,\pm1,\pm2,..., \label{Energy_Landau}%
\end{align}
where $m$ is the magnetic quantum number related to the angular momentum of
the electron. The eigenfunction is also obtained as
\begin{align}
\Psi_{nm}\left(  \rho,\varphi,z\right)   &  =N_{nm}\mathrm{e}^{-\frac{\rho
^{2}}{4l_{B}^{2}}}\left(  \frac{\rho}{l_{B}}\right)  ^{|m|}L_{n}^{|m|}\left(
\frac{\rho^{2}}{2l_{B}^{2}}\right)  \frac{\mathrm{e}^{im\varphi}}{\sqrt{2\pi}%
}\nonumber\label{WF_of_electron}\\
&  =R_{nm}\left(  \rho\right)  \frac{\mathrm{e}^{im\varphi}}{\sqrt{2\pi}},
\end{align}
where $l_{B}=\sqrt{\hbar/eB}$ is the magnetic length, $N_{nm}=\left(
n!/(n+|m|)!\right)  ^{1/2}2^{-|m|/2}l_{B}^{-1}$ is the normalization
constant,\ and $L_{n}^{|m|}(x)$ is the associated Laguerre polynomials. In
this picture, we call $n$ the principal quantum number. Its relation to
ordinary Landau index $N$ is $N=n+(\left\vert m\right\vert +m)/2$. This leads
to $N=n$ for states with $m\leq0$. Each Landau level with given $N$ is
multiply degenerated with respect to $n$ and $m$ due the finite system size
with the degeneracy factor given by $S/(2\pi l_{B}^{2})$, where $S$ is the
area of 2DEG.

\begin{figure}[h]
\centering\includegraphics[scale=0.45]{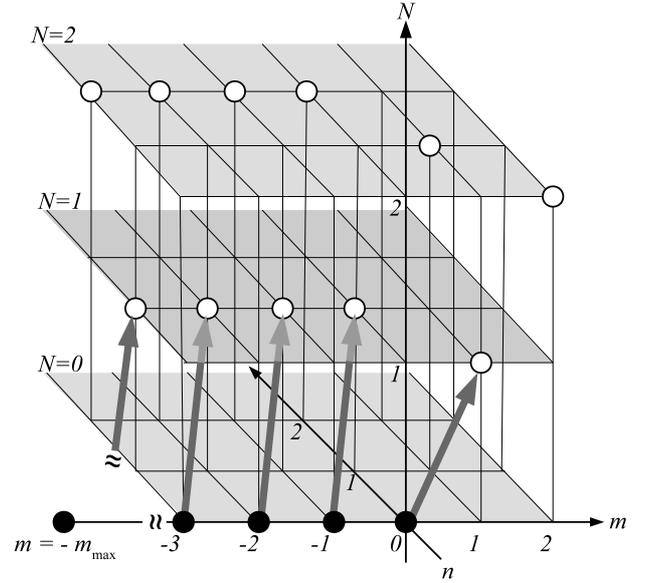}\caption{Allowed
transitions from the lowest LL ($N=0$) to the second LL ($N=1$) are indicated
by the gray thick allows. The opened circles denote unoccupied states, whereas
closed ones are occupied.}%
\label{Degeneracy_Landau_Level}%
\end{figure}

For example, the lowest Landau level (LLL) $N=0$ is obtained by the condition
$n+(\left\vert m\right\vert +m)/2=0$, which leads to $n=0$ and $m\leq0$. The
probability density for the electron with the wave function
(\ref{WF_of_electron}) has the maximal value at $\rho=\sqrt{2\left\vert
m\right\vert +1}l_{B}$. This means that the electron is distributed on the
circle with the radius $\sqrt{2\left\vert m\right\vert +1}l_{B}$. Because the
expectation value of $r^{2}$ is given by $2(\left\vert m\right\vert
+1)l_{B}^{2}$, we find that the electron state covers the area $2\pi l_{B}%
^{2}$. Then, the maximum $m$ for the disk geometry is given
by\cite{Yoshioka2002book}%
\begin{equation}
m_{\text{max}}\simeq\frac{S}{2\pi l_{B}^{2}}=\frac{R^{2}}{2l_{B}^{2}},
\label{Maxima_eAM}%
\end{equation}
which allows us to define the filling factor as%
\begin{equation}
\nu\equiv\frac{N_{e}}{m_{\text{max}}}\simeq2\pi l_{B}^{2}\frac{N_{e}}{\pi
R^{2}},
\end{equation}
where $N_{e}$ is the total number of electrons on the disk. Throughout this
paper, we concentrate on the system with the filling factor $\nu=1$, where the
Fermi energy lies in the gap between the LLL and the second Landau level
(2LL). We display the energy diagram of the axial symmetric 2DEG system as
shown in Fig.\ref{Degeneracy_Landau_Level}. 

\section{Photocurrent Induced by the Optical Vortex}

Here, we investigate the interaction of a Landau-quantized 2DEG with the
Bessel OV by applying the linear response theory. We start with the following
total Hamiltonian, which contains the non-perturbative Hamiltonian
(\ref{Non-perturmativeHamil}) interacting with the vector potential of the
OV:
\begin{equation}
H\mathbf{=}H_{0}+\Delta H=\frac{1}{2m_{e}}\left[  -i\hbar\mathbf{\nabla
}+e\mathbf{A}^{\text{ext}}(\mathbf{r})\right]  ^{2}-\boldsymbol{A}%
_{\ell,\sigma}^{\text{OV}}\cdot\mathbf{j},
\end{equation}
where $\boldsymbol{A}_{\ell,\sigma}^{\text{OV}}$ is given by
Eq.~(\ref{VectorPotentialInPA}), and the electric current is determined by
$\mathbf{j}=\frac{e}{m_{e}}\left(  \mathbf{p}+e\mathbf{A}^{\text{ext}}\right)
$. We neglect the electron spin.

The Kubo formula for $i$-component of the induced current is written as
\cite{Kubo1957,Ando1975}%
\begin{align}
\delta j_{i}\left(  \omega\right)   &  =-\sum_{n,m}\sum_{n^{\prime},m^{\prime
}}\left(  f(E_{n,m})-f(E_{n^{\prime},m^{\prime}})\right) \nonumber\\
&  \times\frac{\langle n,m|j_{i}|n^{\prime},m^{\prime}\rangle\langle
n^{\prime},m^{\prime}|\boldsymbol{A}_{\ell,\sigma}^{\text{OV}}\cdot
\mathbf{j}|n,m\rangle}{E_{n,m}-E_{n^{\prime},m^{\prime}}+\hbar\omega+i\delta}.
\label{Kubo_formula}%
\end{align}
where $f(E_{n,m})$ is the Fermi distribution $f(\epsilon)=\left[  \exp
\beta\left(  \epsilon-\mu\right)  +1\right]  ^{-1}$ with a chemical potential
$\mu$ and an inverse temperature $\beta$, and $|n,m\rangle$ is the electron
wavefunction in Eq.(\ref{WF_of_electron}). From now on, we assume
zero-temperature limit and keeping the chemical potential lie between th LLL
($N=0$) and the second LL ($N=1$).

It should be mentioned that, although we work in the cylindrical coordinates,
which manifest the symmetry of the OV, our final results, of course, are not
specific to a particular coordinate system. Alternatively, we can consider the
spherical coordinates and examine the multipole expansion by the vector
spherical harmonics (VSH) of currents in Eq. (\ref{Kubo_formula}) as discussed
in Appendix~B where we obtain the general expression in Eq.~(\ref{Int_VSH}).
We also show that the selection rules for the dipole transitions in
Eq.~(\ref{Selection_rule_VSH}) are consistent with the results obtained
without multipole expansion in Eq.~(\ref{Selection_rule}).

Let us now return to discussion without multipole expansion. To investigate
the OV-induced photocurrent, we adopt the chiral basis $j_{\pm}(=j_{x}\pm
ij_{y})$.\ First, we consider the matrix element of photocurrent $j_{\pm}$
that can be written as%
\begin{equation}
\langle n,m|j_{\pm}|n^{\prime},m^{\prime}\rangle=i\frac{e}{\hbar}%
d(E_{n,m}-E_{n^{\prime},m^{\prime}})C_{n,m}^{n^{\prime},m^{\prime}}%
\delta_{m^{\prime},m\pm1} \label{Matrix_transition}%
\end{equation}
where $d(\ll R)$ is the thickness of 2DEG and we denote the radial integral as%
\begin{equation}
C_{n,m}^{n^{\prime},m^{\prime}}=\int d\rho\rho^{2}R_{n^{\prime},m^{\prime}%
}(\rho)R_{n,m}(\rho).
\end{equation}
Here, we obtain the selection rule $m^{\prime}=m\pm1$\ from the azimuthal
integral $\int_{0}^{2\pi}\frac{d\varphi}{2\pi}\mathrm{e}^{i(m^{\prime}%
-m\pm1)\varphi}$. After calculating the radial integral $C_{n,m}^{n^{\prime
},m^{\prime}}$ and the energy factor $E_{n,m}-E_{n^{\prime},m^{\prime}}$ by
Eq.(\ref{Energy_Landau}), we can obtain the matrix elements of the
photocurrent as%
\begin{align}
\langle n,m|j_{+}|n^{\prime},m+1\rangle &  =%
\begin{cases}
0\text{ \ \ \ \ \ \ \ \ \ \ \ \ \ \ for\hspace{3pt}}n^{\prime}=n\text{,\hspace
{3pt}}m<0,\\
-iedl_{B}\omega_{c}\sqrt{2n+\left\vert m\right\vert +m+2}\\
\text{ \ \ \ \ \ \ \ \ \ \ \ \ \ \ \ \ \ for\hspace{3pt}}n^{\prime}=n\text{,
}m\geq0,\\
0\text{ \ \ \ \ \ \ \ \ \ \ \ \ \ \ \ for\hspace{3pt}}n^{\prime}%
=n-1\text{,\hspace{3pt}}m\geq0,\\
iedl_{B}\omega_{c}\sqrt{2n+\left\vert m\right\vert +m+2}\\
\text{ \ \ \ \ \ \ \ \ \ \ \ \ \ \ \ \ \ for\hspace{3pt}}n^{\prime
}=n+1\text{,\hspace{3pt}}m<0,\\
0\text{ \ \ \ \ \ \ \ \ \ \ \ \ \ \ \ for otherwise,}%
\end{cases}
,\\
\langle n,m|j_{-}|n^{\prime},m-1\rangle &  =%
\begin{cases}
0\text{ \ \ \ \ \ \ \ \ \ \ \ \ \ \ \ for\hspace{3pt}}n^{\prime}%
=n\text{,\hspace{3pt}}m\leq0,\\
iedl_{B}\omega_{c}\sqrt{2n+\left\vert m\right\vert +m}\\
\text{ \ \ \ \ \ \ \ \ \ \ \ \ \ \ \ \ \ for\hspace{3pt}}n^{\prime
}=n\text{,\hspace{3pt}}m>0,\\
-iedl_{B}\omega_{c}\sqrt{2n+\left\vert m\right\vert +m}\\
\text{ \ \ \ \ \ \ \ \ \ \ \ \ \ \ \ \ \ for\hspace{3pt}}n^{\prime
}=n-1\text{,\hspace{3pt}}m\leq0,\\
0\text{ \ \ \ \ \ \ \ \ \ \ \ \ \ \ \ for\hspace{3pt}}n^{\prime}%
=n+1\text{,\hspace{3pt}}m>0,\\
0\text{ \ \ \ \ \ \ \ \ \ \ \ \ \ \ \ for otherwise,}%
\end{cases}
.
\end{align}

For the filling factor $\nu=1$ $(n=0,m\leq0)$, these matrix elements reduce to%
\begin{align}
\langle0,m|j_{+}|n^{\prime},m+1\rangle &  =%
\begin{cases}
-iedl_{B}\omega_{c}\sqrt{2}\text{ \ \ \ \ \ \ \ for\hspace{3pt}}n^{\prime
}=0\text{,\hspace{3pt}}m=0\\
iedl_{B}\omega_{c}\sqrt{2}\text{ \ \ \ \ \ for\hspace{3pt}}n^{\prime
}=1\text{,\hspace{3pt}}m<0\\
0\text{ \ \ \ \ \ \ \ \ \ \ \ \ \ \ \ \ \ \ \ \ \ \ for otherwise}%
\end{cases}
,\\
\langle0,m|j_{-}|n^{\prime},m-1\rangle &  =0\text{ \ \ for all }n^{\prime},m.
\end{align}
\ \ Therefore, we find that the transition is allowed only\ $N\rightarrow
N+1$\cite{Ando1975}. We summarize possible transitions from the LLL ($N=0$) to
the second LL ($N=1$) as follows,%

\begin{equation}
\left\{
\begin{array}
[c]{cccc}%
\left(  n,m,N\right)  &  & \left(  n^{\prime},m^{\prime},N^{\prime}\right)  &
\\
(0,0,0) & \rightarrow & (0,1,1) & \text{for }m=0\\
(0,m,0) & \rightarrow & (1,m+1,1) & \text{for }m<0
\end{array}
\right.  . \label{Selection_rule}%
\end{equation}

Next, we consider the matrix element of the minimal coupling of 2DEG with OV.
As shown in Appendix A, the dipole approximation is justified in our model.
Then the matrix element for the photon absorption in this approximation is
obtained as%
\begin{align}
\langle n^{\prime},m^{\prime}|\boldsymbol{A}_{\ell,\sigma}^{\text{OV}}%
\cdot\mathbf{j}|n,m\rangle &  \sim i\frac{e}{\hbar}(E_{n^{\prime},m^{\prime}%
}-E_{n,m})\langle n^{\prime},m^{\prime}|\boldsymbol{A}_{\ell,\sigma
}^{\text{OV}}\cdot\mathbf{r}|n,m\rangle\nonumber\\
&  =A_{0}\frac{ed}{\hbar}\sqrt{\frac{k_{\perp}}{4\pi}}(E_{n,m}-E_{n^{\prime
},m^{\prime}})D_{n,m,\ell}^{n^{\prime},m^{\prime}}\nonumber\\
&  \times\delta_{m^{\prime},m+\ell+\sigma},\label{Matrix_transition2}%
\end{align}
where we denoted the radial integral as
\begin{equation}
D_{n,m,\ell}^{n^{\prime},m^{\prime}}=\int d\rho\rho^{2}R_{n^{\prime}%
,m^{\prime}}(\rho)R_{n,m}(\rho)J_{\ell}(k_{\perp}\rho).
\end{equation}
We also obtain the selection rule $m^{\prime}=m+\ell+\sigma$ from the
azimuthal integral $\int_{0}^{2\pi}\frac{d\varphi}{2\pi}\mathrm{e}%
^{i(m-m^{\prime}+\ell+\sigma)\varphi}$, where $\sigma=\pm1$. This means that
the OV can transfer its TAM to the electron via the dipole interaction.

We note that, when we fix the filling factor $\nu=1$ (the chemical potential
lies between $N=0$ and\ $N=1$), the left-handed current is not induced.
Therefore, only the right-handed current arises by transferring the optical
TAM, $J=1$. Because the OV carries the SAM $\sigma=\pm1$, the OAM and SAM must
be $\ell=0$, $\sigma=1$, or $\ell=2$, $\sigma=-1$, respectively, with the
other transitions being prohibited.

On the other hand, if we apply the external magnetic field anti-parallel to
the light traveling, since it corresponds to the time inverse, the electron in
the LLL carries positive value angular momentum. Then, to excite the electron
in the LLL, the electron can absorb the optical TAM $J=-1$. As a result, the
possible absorptions are reduced to $\ell=0$, $\sigma=-1$, and $\ell=-2$,
$\sigma=1$.

Next, we calculate the photocurrent using the Kubo formula. For the transition
from $N=0$ to $N=1$ with $\nu=1$, the OV-induced current~(\ref{Kubo_formula})
reduces to%
\begin{equation}
\delta j_{\ell}^{+}\left(  \omega,B\right)  =-i\frac{F^{\ell}\left(  B\right)
}{\hbar\omega-\hbar\omega_{c}+i\delta}, \label{Kubo_formula2}%
\end{equation}
where $\ell=0$ or $2$ and the factors $F^{\ell}$ are given by
\begin{equation}
F^{\ell}\left(  B\right)  =A_{1}C_{0,0}^{0,1}D_{0,0,\ell}^{0,1}+A_{1}%
\sum_{m<0}^{-m_{\max}}C_{0,m}^{1,m+1}D_{0,m,\ell}^{1,m+1}, \label{I02}%
\end{equation}
with $A_{1}=A_{0}e^{2}d^{2}\omega_{c}^{2}\sqrt{k_{\perp}/4\pi}/V$. In the
summation with respect to $m$, by using the explicit form, $L_{1}^{k}\left(
x\right)  =1+k-x$, only one term corresponding to an edge current along the
circle with the radius $R$ survives. The other terms corresponding to the bulk
currents cancel each other. After some algebra, we obtain%
\begin{align}
F^{\ell}\left(  B\right)   &  \sim\frac{F_{0}}{\sqrt{\alpha^{5}}}\left(
\frac{1+\alpha^{2}}{\alpha^{2}}\frac{\Phi_{0}^{2}}{\lambda_{e}^{2}R^{2}B^{2}%
}\left[  1+\frac{\Phi_{0}}{2\pi R^{2}B}\right]  \mathrm{e}\right)  ^{\frac{\pi
R^{2}}{\Phi_{0}}B}\nonumber\\
&  \times\int_{0}^{k_{\perp}R}dxx^{2m_{\text{max}}\left(  B\right)
+3}\mathrm{e}^{-\frac{x^{2}}{2k_{\perp}^{2}l_{B}^{2}}}J_{\ell}(x),
\label{Final expression}%
\end{align}
where $\ell$ is $0$\ or $2$, $F_{0}=A_{0}e^{2}d^{2}c^{2}/V\sqrt{2\pi
\lambda_{e}\mathrm{e}}$, $x=k_{\perp}\rho$, which has an order of magnitude of
unity. $\Phi_{0}=2\pi\hbar/e$ is the flux quantum, and $\lambda_{e}=2\pi
\hbar/m_{e}c$ is the electron Compton wavelength.

\section{Physical Meaning of Cancellation of Bulk Currents}

\begin{figure}[h]
\centering\includegraphics[scale=0.45]{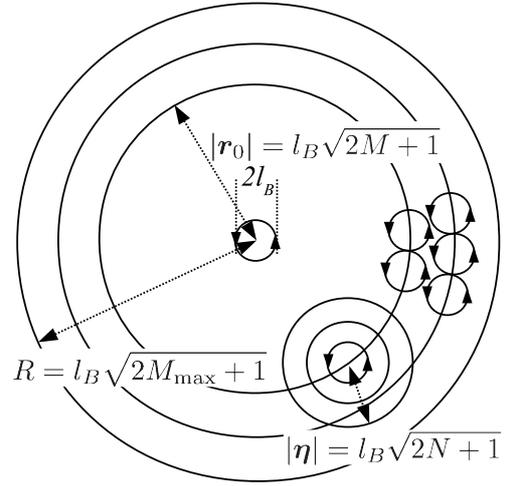}\caption{Schematics
of the classical orbits of LLs. The quantum number $M$ assigns the radius of
the guiding center $r_{0}$, whereas $N$ is the radius of Larmor orbit $\eta$.}%
\label{Larmor_orbits}%
\end{figure}

In this section, we present a physical interpretation on the reason why the
bulk currents are cancelled out, based on the coherent state representation.
Introducing the Larmor radius vector $\boldsymbol{\eta}$ and the guiding
center vector $\boldsymbol{r}_{0}=\left(  x_{0},y_{0}\right)  \ $%
satisfying$\ \left(  \eta_{x},\eta_{y}\right)  =\left(  x-x_{0},y-y_{0}%
\right)  $, we can rewrite the 2DEG Hamiltonian as%
\begin{equation}
H_{0}=\frac{1}{2}m_{e}\omega_{c}^{2}\left(  \eta_{x}^{2}+\eta_{y}^{2}\right)
=\frac{1}{2}m_{e}v_{\perp}^{2},
\end{equation}
where we note the relation $\left\vert \boldsymbol{\eta}\right\vert =v_{\perp
}/\omega_{c}$. We can then define the non-commuting operators which satisfy%
\begin{equation}
\left[  \hat{\eta}_{x},\hat{\eta}_{y}\right]  =-il_{B},\ \left[  \hat{x}%
_{0},\hat{y}_{0}\right]  =il_{B}.
\end{equation}
We find that one electron occupies the area determined by the uncertainty
principle:%
\begin{equation}
\Delta S=\Delta x_{0}\Delta y_{0}=2\pi l_{B}^{2}. \label{Uncertainty}%
\end{equation}
Then we can define the ladder operators%
\begin{align}
a  &  =\frac{1}{\sqrt{2}l_{B}}\left(  \hat{\eta}_{x}-i\hat{\eta}_{y}\right)
,\ a^{\dag}=\frac{1}{\sqrt{2}l_{B}}\left(  \hat{\eta}_{x}+i\hat{\eta}%
_{y}\right)  ,\nonumber\\
b  &  =\frac{1}{\sqrt{2}l_{B}}\left(  \hat{x}_{0}+i\hat{y}_{0}\right)
,\ b^{\dag}=\frac{1}{\sqrt{2}l_{B}}\left(  \hat{x}_{0}-i\hat{y}_{0}\right)  ,
\end{align}
with $\left[  a,a^{\dag}\right]  =\left[  b,b^{\dag}\right]  =1$, and $\left[
a,b^{(\dag)}\right]  =0$. The eigenstates are thus determined by the two
integer quantum numbers, $N$ and $M$, associated with the two ladder
operators,%
\begin{align}
a^{\dag}|N,M\rangle &  =\sqrt{N+1}|N+1,M\rangle,\text{ \ }a|N,M\rangle
=\sqrt{N}|N-1,M\rangle\nonumber\\
&  \text{\ \ \ \ \ \ \ \ \ \ \ \ \ \ \ for }N>0;\\
b^{\dag}|N,M\rangle &  =\sqrt{M+1}|N,M+1\rangle,\text{ \ }b|N,M\rangle
=\sqrt{M}|N,M-1\rangle\nonumber\\
&  \text{\ \ \ \ \ \ \ \ \ \ \ \ \ \ \ for }M>0.
\end{align}
Then in terms of the two ladder operators, the Hamiltonian and the angular
momentum operator are written by
\begin{equation}
H_{0}=\hbar\omega_{c}\left(  a^{\dag}a+\frac{1}{2}\right)  =\hbar\omega
_{c}\left(  N+\frac{1}{2}\right)  \label{Hamiltonian2}%
\end{equation}%
\begin{equation}
L_{z}=\hbar\left(  a^{\dag}a-b^{\dag}b\right)  =\hbar\left(  N-M\right)  .
\end{equation}
Here, comparing above eigenvalues with Eq.(\ref{Energy_Landau})\ and
$L_{z}=\hbar m$, we can determine the relation between $n$, $m$ and $N$, $M$
as%
\begin{equation}
N=n+\frac{\left\vert m\right\vert +m}{2},\ M=n+\frac{\left\vert m\right\vert
-m}{2}.
\end{equation}
The average value of the guiding center operator $\boldsymbol{\hat{r}}_{0}$
gives%
\begin{equation}
\left\langle N,M\left\vert \boldsymbol{\hat{r}}_{0}\right\vert
N,M\right\rangle =0,
\end{equation}
but its absolute value leads to%
\begin{equation}
\left\langle \left\vert \boldsymbol{r}_{0}\right\vert \right\rangle
_{N,M}\equiv\sqrt{\left\langle N,M\left\vert \boldsymbol{\hat{r}}_{0}%
^{2}\right\vert N,M\right\rangle }=l_{B}\sqrt{2M+1}.
\end{equation}
Similarly, the average value of Larmor radius operator gives%
\begin{equation}
\left\langle N,M\left\vert \boldsymbol{\hat{\eta}}\right\vert N,M\right\rangle
=0,
\end{equation}
but its absolute value is given by%
\begin{equation}
\left\langle \left\vert \boldsymbol{\eta}\right\vert \right\rangle
_{N,M}\equiv\sqrt{\left\langle N,M\left\vert \boldsymbol{\hat{\eta}}%
^{2}\right\vert N,M\right\rangle }=l_{B}\sqrt{2N+1}.
\end{equation}
Therefore, the arbitrary state $|N,M\rangle$ distributes at the center of the
Larmor motion with radius $\left\langle \left\vert \boldsymbol{\eta
}\right\vert \right\rangle _{N,M}=l_{B}\sqrt{2N+1}$\ is located at the
position of guiding center
\begin{equation}
\left\langle \left\vert \boldsymbol{r}_{0}\right\vert \right\rangle
_{N,M}=l_{B}\sqrt{2M+1}. \label{Guiding_center}%
\end{equation}
\ The geometric meaning of\ this is illustrated in Fig. \ref{Larmor_orbits}.
When we focus on the LLL, that is, $n=0$ and$\ m\leq0$, we see $N=0$ and
$M=\left\vert m\right\vert $. Therefore, the guiding center in the LLL is
$\left\langle \left\vert \boldsymbol{r}_{0}\right\vert \right\rangle
_{0,\left\vert m\right\vert }=l_{B}\sqrt{2\left\vert m\right\vert +1}$, and
the Larmor radius in it is $\left\langle \left\vert \boldsymbol{\eta
}\right\vert \right\rangle _{0,\left\vert m\right\vert }=l_{B}$%
.

This kind of distribution can semi-classically be described by the coherent
states. We now introduce the displacement operators%
\begin{equation}
D\left(  x_{0},y_{0}\right)  =e^{-i\left(  x_{0}\hat{y}_{0}-y_{0}\hat{x}%
_{0}\right)  /l_{B}^{2}},\ D\left(  \eta_{x},\eta_{y}\right)  =e^{i\left(
\eta_{x}\hat{\eta}_{y}-\eta_{y}\hat{\eta}_{x}\right)  /l_{B}^{2}}.
\end{equation}
The first displacement operator\ $D\left(  x_{0},y_{0}\right)  $\ generates a
displacement to the position at the guiding center $\left\vert \boldsymbol{r}%
_{0}\right\vert =\sqrt{x_{0}^{2}+y_{0}^{2}}$. Since $D\left(  x_{0}%
,y_{0}\right)  $ commutes with the Hamiltonian $H_{0}$,\ the guiding center is
the constant of motion. Therefore, the Hamiltonian $H_{0}$\ does not depend on
quantum number $M$. On the other hand, the second displacement operator
$D\left(  \eta_{x},\eta_{y}\right)  $ generates a displacement to the position
$\left\vert \boldsymbol{\eta}\right\vert =\sqrt{\eta_{x}^{2}+\eta_{y}^{2}}%
$.\ Since $D\left(  \eta_{x},\eta_{y}\right)  $ does not commute with the
Hamiltonian $H_{0}$,\ the coherent state is not an eigenstate of the Hamiltonian.

Applying these displacement operators to the ground state $|0,0\rangle
=|n_{a}=0,n_{b}=0\rangle$,\ we can thus construct the coherent state,%
\begin{align}
|X_{0},Y_{0};\eta_{x},\eta_{y}\rangle &  =D\left(  \eta_{x},\eta_{y}\right)
D\left(  x_{0},y_{0}\right)  |0,0\rangle\nonumber\\
&  =e^{-\left(  \left\vert \alpha\right\vert ^{2}+\left\vert \beta\right\vert
^{2}\right)  /2}e^{\alpha a^{\dag}}e^{\beta b^{\dag}}|0,0\rangle,
\label{Coherent_state}%
\end{align}
where $\alpha$ and $\beta$ are eigenvalues of the annihilation operators $a$
and $b$ of the eigenstate $|x_{0},y_{0};\eta_{x},\eta_{y}\rangle$. That is,
these eigenstates satisfy%
\begin{align}
a|x_{0},y_{0};\eta_{x},\eta_{y}\rangle &  =\frac{\eta_{x}-i\eta_{y}}{\sqrt
{2}l_{B}}|x_{0},y_{0};\eta_{x},\eta_{y}\rangle\nonumber\\
&  =\frac{\left\vert \boldsymbol{\eta}\right\vert e^{-i\varphi}}{\sqrt{2}%
l_{B}}|x_{0},y_{0};\eta_{x},\eta_{y}\rangle,\\
b|x_{0},y_{0};\eta_{x},\eta_{x}\rangle &  =\frac{x_{0}+iy_{0}}{\sqrt{2}l_{B}%
}|x_{0},y_{0};\eta_{x},\eta_{y}\rangle\nonumber\\
&  =\frac{\left\vert \boldsymbol{r}_{0}\right\vert e^{i\varphi}}{\sqrt{2}%
l_{B}}|x_{0},y_{0};\eta_{x},\eta_{y}\rangle,
\end{align}
and the eigenvalues $\alpha$ and $\beta$ are given by%
\begin{align}
\alpha &  =\frac{\eta_{x}-i\eta_{y}}{\sqrt{2}l_{B}}=\frac{\left\vert
\boldsymbol{\eta}\right\vert e^{-i\varphi}}{\sqrt{2}l_{B}},\\
\beta &  =\frac{x_{0}+iy_{0}}{\sqrt{2}l_{B}}=\frac{\left\vert \boldsymbol{r}%
_{0}\right\vert e^{i\varphi}}{\sqrt{2}l_{B}}.
\end{align}

To see the absence of bulk currents, we pay our attention to one coherent
state at the guiding center $\boldsymbol{r}_{0}$, which produces the circular
current by the Larmor motion with radius $\left\vert \boldsymbol{\eta
}\right\vert $. Because of the uncertainty (\ref{Uncertainty}), it seems that
the circular current flows the edge of the area $\Delta S$. When the LLs state
can be constructed by the superposition of the coherent states, the
superposition produces\ contact points of the circular current at the center
$\boldsymbol{r}_{0}$ with the surrounding circular currents. Thus, the
circular current at the center $\boldsymbol{r}_{0}$\ is canceled by the
surrounding circular current.\ Such the cancellation occurs on whole system
except to the edge,\ we can say the bulk currents are all cancelled out,
i.e.,
\begin{equation}
\boldsymbol{j}_{\text{bulk}}=0. \label{Bulk_cuurent_cancel}%
\end{equation}

\section{Magnetization Induced by Edge Current}

Now, we naturally expect that the edge currents induce an orbital
magnetization, which can be observed experimentally. The magnetic vector
potential at the position $r$ induced by the magnetization at the guiding
center $\boldsymbol{M}\left(  \boldsymbol{r}_{\boldsymbol{0}}\right)  $ is
given by%
\begin{equation}
\boldsymbol{A}\left(  \boldsymbol{r}\right)  =\frac{\mu_{0}}{4\pi}\int
_{D}\frac{\boldsymbol{\nabla}_{\boldsymbol{r}_{0}}\times\boldsymbol{M}\left(
\boldsymbol{r}_{0}\right)  }{\left\vert \boldsymbol{r}-\boldsymbol{r}%
_{0}\right\vert }dV_{0}+\frac{\mu_{0}}{4\pi}\int_{\partial D}\frac
{\boldsymbol{M}\left(  \boldsymbol{r}_{0}\right)  \times\boldsymbol{\hat{n}}%
}{\left\vert \boldsymbol{r}-\boldsymbol{r}_{0}\right\vert }dS_{0},
\label{Vector_potential_Mag}%
\end{equation}
where $\partial D$\ represents the edge of the 2D system $D$, and
$\boldsymbol{\hat{n}}$ is a normal vector with respect to the edge $\partial
D$, $dS_{0}$ and $dV_{0}$ indicate that the integration is done with respect
to a variable $\boldsymbol{r}_{0}$. The first term can be regarded as the
vector potential induced by the bulk current, $\boldsymbol{j}_{\text{bulk}%
}=\boldsymbol{\nabla}_{\boldsymbol{r}_{0}}\times\boldsymbol{M}\left(
\boldsymbol{r}_{0}\right)  $, whereas the second term is due to the edge
current at the system size $R$,%
\begin{equation}
\boldsymbol{j}_{\text{edge}}=\boldsymbol{M}\left(  R\right)  \times
\boldsymbol{\hat{n}}. \label{Edge_magnetization_current}%
\end{equation}
However, since the bulk currents cancel out by Eq. (\ref{Bulk_cuurent_cancel})
as mentioned in the previous section, only the edge current contributes to the
magnetization in Eq.~(\ref{Vector_potential_Mag}).

In Eq. (\ref{Edge_magnetization_current}), since the normal vector with
respect to the edge of circular disk is given by $\boldsymbol{\hat{n}%
}=\boldsymbol{\hat{e}}_{\rho}$, and the edge current flows along the edge,
$\boldsymbol{j}_{\text{edge}}\sim\delta j_{\ell}^{+}\left(  \omega,B\right)
\boldsymbol{\hat{e}}_{\varphi}$, the magnetization points along the
$z$-direction, $\boldsymbol{M}\left(  R\right)  \sim\mathcal{M}_{\ell}\left(
\omega,B\right)  \boldsymbol{\hat{e}}_{z}$, where
\begin{equation}
\mathcal{M}_{\ell}\left(  \omega,B\right)  =\delta j_{\ell}^{+}\left(
\omega,B\right)  . \label{eq60}%
\end{equation}
The magnetization in Eq.~(\ref{eq60}) can be regarded as a manifestation of
the magneto-electric effect, since it is induced by the electric field of OV.

The magnetization obviously depends on the external magnetic field. Here, we
imply that the frequency of the OV is always kept in resonance with the
transition from the LLL to 2LL, so that when we apply the external magnetic
field $B$, the excitation energy from the LLL to 2LL is given by $\hbar
\omega_{c}=\hbar eB/m_{e}\sim1.14\times10^{-4}B$[T]eV. To make the transitions
possible, the wavelength of OV must be controlled to satisfy the energy
conservation, $h/\lambda_{\text{OV}}c=\hbar\omega_{c}$. Then the wavelength of
OV and wavenumber should be $\lambda_{\text{OV}}=2\pi c/\omega_{c}%
\sim1.07\times10^{-2}B^{-1}$[T$^{-1}$]m and $k=2\pi/\lambda_{\text{OV}}%
\sim5.87\times10^{2}B$[T]m$^{-1}$, respectively. As a consequence, when the
magnetic field increases, the transverse wavenumber $k_{\perp}$ should be
increased to hold the ratio, $\alpha=k_{\perp}/k_{z}$, according to the
following expression%
\begin{equation}
k_{\perp}r_{\perp}^{\ell,i}=\frac{\alpha}{\sqrt{1+\alpha^{2}}}\frac
{\lambda_{e}Br_{\perp}^{\ell,i}}{\Phi_{0}}\simeq587\alpha Br_{\perp}^{\ell,i},
\end{equation}
which leads to shrinking the dark ring radius of the OV, see
Eq.~(\ref{Dark_ring_radius}). We also assume that the chemical potential $\mu$
is between the LLL and the second LL.

\begin{figure}[ptb]
\centering\includegraphics[scale=0.65]{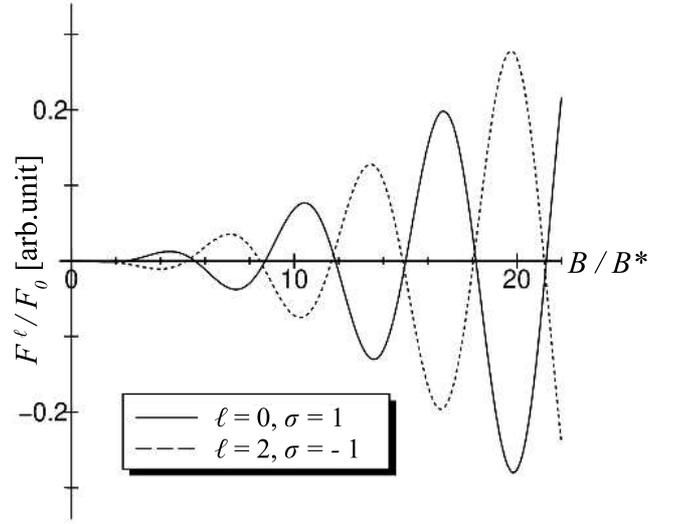}\caption{Magnetic
field dependence of $F^{\ell}$ with $R=10^{-2}$~m and $\alpha=0.1$, when the
chemical potential is kept\ between the LLL and the second LL. The solid line
is\ for $\ell=0$ and the dotted line $\ell=2$. We scale the horizontal axis by
$B^{\ast}=\Phi_{0}/\alpha\lambda_{e}R=1.70\times10^{-3}/\alpha R$ [T].}%
\label{intensity_plot}%
\end{figure}

Figure~\ref{intensity_plot} demonstrates the magnetic field dependencies of
$F^{\ell}$ (which are proportional to the orbital magnetization) for
$\ell=0,2$. Here we introduced the characteristic magnetic field strength,
$B^{\ast}\equiv\Phi_{0}/\alpha\lambda_{e}R$, which corresponds to $k_{\perp
}R\sim1$, and chose $R=10^{-2}$~m and $\alpha=0.1$. Because the radial profile
of OV has the oscillating behavior, the amplitudes of absorption $F^{\ell}$
oscillate with increasing the magnetic field strength, and have vanishing
points. As discussed in Ref. \citen{Takahashi2018}, when $r_{\perp}^{\ell
,i}=R$,\ the roots of Bessel function, $J_{\ell}(k_{\perp}\left[  B\right]
R)=0$, cause the vanishing points of absorption. Physically it means that when
the dark rings of OV coincide with the peak of electron distribution on the
system edge, the orbital magnetization disappears. It is significant that this
disappearance is induced despite non-zero total intensity, which is related to
the fact that the photocurrent flows only along the edge. It is worth a mention that the similar result is obtained by using
the cylindrical vector beams.\cite{FujitaSato2018}

\section{Concluding Remarks}

We discussed Landau level spectroscopy of a two-dimensional electron gas with
modified selection rules illuminated by the optical vortex beam which carries
orbital angular momentum in the paraxial approximation. The absorption of the
vortex beam occurs for $\sigma=1$ (positive helicity) and $\ell=0$ or
$\sigma=-1$ (negative helicity) and $\ell=2$. We reconstructed the
minimal-coupling Hamiltonian by expanding by the vector spherical Harmonics
and confirmed that the dipole transitions are allowed when the optical beam
carries the total angular momentum $J=1$, $0$, or $-1$. This result is
consistent with calculation without multipolar expansion in our previous
Letter \cite{Takahashi2018}.

In the framework of Kubo's linear response theory, we found that the
absorption of optical vortex induces the photocurrents, which flow only along
the edge of the system. The cancellation of the bulk currents was interpreted
in terms of the coherent state representation. Consequently, we demonstrated
how the orbital magnetization is induced by the edge currents.

\begin{acknowledgments}
We thank K. Oto, Y. Yamada, N. Yokoshi, S. Hashiyada, J. Goryo, and Y. Togawa for fruitful discussions. This work was supported by the Chirality Research Center in Hiroshima
University, JSPS KAKENHI Grant 25220803, 17H02923, the JSPS Core-to-Core
Program, A. Advanced Research Networks, and the JSPS Bilateral (Japan-Russia)
Joint Research Projects. I.P. acknowledges financial support by Ministry of Education and Science of the Russian Federation, Grant No. MK-1731.2018.2 and by Russian Foundation for Basic Research (RFBR), Grant 18-32-00769(mol\_a).
\end{acknowledgments}

\appendix

\section{Dipole Approximation in Minimal Coupling}

We present the derivation of Eq. (\ref{Matrix_transition2}).\ We need to
compute commutator $\left[  H_{0},\mathbf{A}^{\text{OV}}\right]  $.\ After
some calculations, we obtain%
\begin{equation}
\lbrack H_{0},\mathbf{A}^{\text{OV}}]\!=\!\mathbf{A}^{\text{OV}}\!\left\{
\frac{\hbar^{2}k_{\text{OV}}^{2}}{2m_{e}}\!-\!\left(  \frac{\hbar
\mathbf{k}_{\text{OV}}}{m_{e}}\right)  \!\cdot\!(-i\hbar\mathbf{\nabla
})\!+\!\frac{e}{m_{e}}\mathbf{A}^{\text{ext}}\!\cdot\!\hbar\mathbf{k}%
_{\text{OV}})\right\}  , \label{Comm_Rel}%
\end{equation}
where $\mathbf{k}_{\text{OV}}$ is the wavenumber vector of OV and
$k_{\text{OV}}$ is its magnitude.

Next, we evaluate the matrix element of the minimal coupling term $\langle
n^{\prime},m^{\prime}|\mathbf{A}^{\text{OV}}\cdot\mathbf{j}|n,m\rangle$ with a
current operator, $\mathbf{j}=\frac{e}{m_{e}}\left(  \mathbf{p}+e\mathbf{A}%
^{\text{ext}}\right)  $. Noting that this current operator satisfies
$\mathbf{p}+e\mathbf{A}^{\text{ext}}=\frac{im_{e}}{\hbar}[H_{0},\mathbf{r}]$,
by using the commutation relation (\ref{Comm_Rel}), we obtain%
\begin{align}
\langle n^{\prime},m^{\prime}|\mathbf{A}^{\text{OV}}\cdot\mathbf{j}|n,m\rangle
&  =\frac{e}{m_{e}}\langle n^{\prime},m^{\prime}|\mathbf{A}^{\text{OV}}%
\cdot\frac{im_{e}}{\hbar}[H_{0},\mathbf{r}]|n,m\rangle\nonumber\\
&  =\frac{ie}{\hbar}(E_{n^{\prime},m^{\prime}}-E_{n,m})\langle n^{\prime
},m^{\prime}|\mathbf{A}^{\text{OV}}\cdot\mathbf{r}|n,m\rangle\nonumber\\
&  -\frac{ie}{\hbar}\frac{\hbar^{2}k_{\text{OV}}^{2}}{2m_{e}}\langle
n^{\prime},m^{\prime}|\mathbf{A}^{\text{OV}}\cdot\mathbf{r}|n,m\rangle
\nonumber\\
&  -\frac{ie}{\hbar}\frac{\hbar e}{m_{e}}\langle n^{\prime},m^{\prime
}|(\mathbf{A}^{\text{ext}}\cdot\mathbf{k}_{\text{OV}})(\mathbf{A}^{\text{OV}%
}\cdot\mathbf{r})|n,m\rangle\nonumber\\
&  -\frac{e}{\hbar}\frac{\hbar^{2}}{2m_{e}}\langle n^{\prime},m^{\prime
}|\mathbf{A}^{\text{OV}}\cdot\mathbf{k}_{\text{OV}}|n,m\rangle\nonumber\\
&  -\frac{e}{\hbar}\frac{\hbar^{2}}{2m_{e}}\langle n^{\prime},m^{\prime
}|(\mathbf{A}^{\text{OV}}\cdot\mathbf{r})\mathbf{k}_{\text{OV}}\cdot
\operatorname{grad}|n,m\rangle. \label{Minimal_coupling_transition}%
\end{align}
The second term is $10^{-11}$ times smaller than the first term for $B=10$T
and can be dropped. Furthermore, the OV travels along $z$-axis and the
wavenumber vector of the OV is approximately described as, $\boldsymbol{k}%
_{\text{OV}}\sim k_{z}\boldsymbol{\hat{e}}_{z}$,\ in the paraxial
approximation. On the other hand, $\mathbf{A}^{\text{ext}},$ $\mathbf{A}%
^{\text{OV}},$\ and, $\operatorname{grad}\Psi_{nm}\left(  \rho,\varphi
,z\right)  $ have no $z$-component. The inner products with $\boldsymbol{k}%
_{\text{OV}}$ in the third, fourth, and fifth terms in Eq.
(\ref{Minimal_coupling_transition}) thus vanishes. As a consequence, the first
term only survives in the matrix element of the minimal coupling,%
\begin{equation}
\langle n^{\prime},m^{\prime}|\mathbf{A}^{\text{OV}}\cdot\mathbf{j}%
|n,m\rangle\sim\frac{ie}{\hbar}(E_{n^{\prime},m^{\prime}}-E_{n,m})\langle
n^{\prime},m^{\prime}|\mathbf{A}^{\text{OV}}\cdot\mathbf{r}|n,m\rangle.
\label{Dipole_matrix_element}%
\end{equation}
which is Eq. (\ref{Matrix_transition2}).

\section{Expansion of Current Operator using Vector Spherical Harmonics}

In the main text, we used the orthogonal basis to represent the current
operator. We here demonstrate that the selection rules derived in the text can
be more directly understood by using the vector spherical harmonics
(VSH)\cite{Barrera1985} as the basis for the current operator.

First, we give the definition of the VSH as the followings,%
\begin{align}
\boldsymbol{Y}_{\ell m}\left(  \theta,\varphi\right)   &  =Y_{\ell m}\left(
\theta,\varphi\right)  \boldsymbol{\hat{e}}_{r},\nonumber\\
\boldsymbol{\Psi}_{\ell m}\left(  \theta,\varphi\right)   &
=r\boldsymbol{\nabla}Y_{\ell m}\left(  \theta,\varphi\right)  ,\\
\boldsymbol{\Phi}_{\ell m}\left(  \theta,\varphi\right)   &  =\boldsymbol{r}%
\times\boldsymbol{\nabla}Y_{\ell m}\left(  \theta,\varphi\right)  ,\nonumber
\end{align}
with a spherical harmonics $Y_{\ell m}\left(  \theta,\varphi\right)  $. Then
the current can be expanded by VSH as%
\begin{align}
\boldsymbol{j}\left(  \boldsymbol{r}\right)   &  =\sum_{\ell=0}^{\infty}%
\sum_{m=-\ell}^{\ell}\left[  j_{\ell m}^{\left(  r\right)  }\left(  r\right)
\boldsymbol{Y}_{\ell m}\left(  \theta,\varphi\right)  +j_{\ell m}^{\left(
1\right)  }\left(  r\right)  \boldsymbol{\Psi}_{\ell m}\left(  \theta
,\varphi\right)  \right. \nonumber\\
&  \left.  \text{ \ \ \ \ \ \ \ \ \ \ }+j_{\ell m}^{\left(  2\right)  }\left(
r\right)  \boldsymbol{\Phi}_{\ell m}\left(  \theta,\varphi\right)  \right]  ,
\end{align}
where we introduced the multipole coefficients $j_{\ell m}^{\left(  r\right)
}\left(  r\right)  $, $j_{\ell m}^{\left(  1\right)  }\left(  r\right)  $, and
$j_{\ell m}^{\left(  2\right)  }\left(  r\right)  $, and spherical coordinates
taken as Fig. \ref{SetupConfig}. Next, the vector potential of OV can also be
expressed in terms of the spherical coordinates,%
\begin{equation}
\boldsymbol{A}_{\ell,\sigma}^{\text{OV}}\left(  \boldsymbol{r}\right)
=\boldsymbol{\eta}_{\sigma}\sqrt{\frac{k_{\perp}}{2\pi}}\sum_{\ell^{\prime
}=-\infty}^{\infty}i^{\ell^{\prime}-\sigma}J_{\ell}\left(  k_{\perp}%
r\sin\theta\right)  J_{\ell^{\prime}}(k_{z}r)e^{i\ell\varphi}e^{i\ell^{\prime
}\theta}.
\end{equation}
where the polarization vector is%
\[
\boldsymbol{\eta}_{\sigma}=-\sigma\frac{\sin\theta}{\sqrt{2}}e^{i\sigma
\varphi}\boldsymbol{\hat{e}}_{r}-\sigma\frac{\cos\theta}{\sqrt{2}}%
e^{i\sigma\varphi}\boldsymbol{\hat{e}}_{\theta}-\frac{i}{\sqrt{2}}%
e^{i\sigma\varphi}\boldsymbol{\hat{e}}_{\varphi}\text{ \ for }\sigma=\pm1.
\]
and we applied a plane wave expansion%
\begin{equation}
e^{ik_{z}r\cos\theta}={\sum\limits_{\ell=-\infty}^{\infty}}i^{\ell}J_{\ell
}(k_{z}r)e^{i\ell\theta}.
\end{equation}
We consider the interaction of the current with the OV as a minimal coupling.
The Hamiltonian is given by%
\begin{align}
H_{\text{int}}  &  =-\int\boldsymbol{j}\left(  \boldsymbol{r}\right)
\cdot\boldsymbol{A}_{\ell,\sigma}^{\text{OV}}\left(  \boldsymbol{r}\right)
d^{3}\boldsymbol{r}\nonumber\\
&  =k_{\perp}^{1/2}\sum_{\ell^{\prime}=-\infty}^{\infty}\sum_{\ell
^{\prime\prime}=0}^{\infty}\sum_{m^{\prime\prime}=-\ell^{\prime\prime}}%
^{\ell^{\prime\prime}}\delta_{m^{\prime\prime},-\left(  \ell+\sigma\right)
}i^{\ell^{\prime}-\sigma}\nonumber\\
&  \times\sqrt{\frac{2\ell^{\prime\prime}+1}{4}\frac{\left(  \ell
^{\prime\prime}-m^{\prime\prime}\right)  !}{\left(  \ell^{\prime\prime
}+m^{\prime\prime}\right)  !}}\nonumber\\
&  \times\int drd\theta\text{ }r^{2}J_{\ell^{\prime}}(k_{z}r)J_{\ell}\left(
k_{\perp}r\sin\theta\right)  e^{i\ell^{\prime}\theta}\nonumber\\
&  \times\left[  \sigma j_{\ell^{\prime\prime}m^{\prime\prime}}^{\left(
r\right)  }\left(  r\right)  \sin^{2}\theta P_{\ell^{\prime\prime}}%
^{m^{\prime\prime}}\left(  \cos\theta\right)  \right. \nonumber\\
&  \left.  \text{ \ \ }+\sigma j_{\ell^{\prime\prime}m^{\prime\prime}%
}^{\left(  1\right)  }\left(  r\right)  \cos\theta\sin\theta\frac{\partial
P_{\ell^{\prime\prime}}^{m^{\prime\prime}}\left(  \cos\theta\right)
}{\partial\theta}\right. \nonumber\\
&  \left.  \text{ \ \ }-m^{\prime\prime}j_{\ell^{\prime\prime}m^{\prime\prime
}}^{\left(  1\right)  }\left(  r\right)  P_{\ell^{\prime\prime}}%
^{m^{\prime\prime}}\left(  \cos\theta\right)  \right. \nonumber\\
&  \left.  \text{ \ \ }-i\sigma m^{\prime\prime}j_{\ell^{\prime\prime
}m^{\prime\prime}}^{\left(  2\right)  }\left(  r\right)  \cos\theta
P_{\ell^{\prime\prime}}^{m^{\prime\prime}}\left(  \cos\theta\right)  \right.
\nonumber\\
&  \left.  \text{ \ \ }+ij_{\ell^{\prime\prime}m^{\prime\prime}}^{\left(
2\right)  }\left(  r\right)  \sin\theta\frac{\partial P_{\ell^{\prime\prime}%
}^{m^{\prime\prime}}\left(  \cos\theta\right)  }{\partial\theta}\right]  .
\label{Int_VSH}%
\end{align}
We here note that the angular momentum conservation $\delta_{m^{\prime\prime
},-\left(  \ell+\sigma\right)  }$\ is provided by integral with respect to the
azimuthal angle $\varphi$.

Considering the dipole transitions, we focus on the dipole moment of
$\boldsymbol{j}\left(  \boldsymbol{r}\right)  $, which corresponds
$\ell^{\prime\prime}=1$. When the current interacts with OV near the optical
axis, we use the limit $k_{\perp}r\sin\theta\ll1$ and apply the formulae
$J_{-\ell}\left(  k_{\perp}r\sin\theta\right)  =\left(  -1\right)  ^{\ell
}J_{\ell}\left(  k_{\perp}r\sin\theta\right)  $, and $J_{\ell}\left(
k_{\perp}r\sin\theta\right)  \sim\left(  k_{\perp}r\sin\theta/2\right)
^{\ell}/\ell!$. We thus arrive at six types of allowed transitions as follows:%
\begin{align}
H_{\text{int}\left(  \ell=0,\sigma=1\right)  }^{\text{dip}}  &  =-ik_{\perp
}^{1/2}\frac{\sqrt{6}}{3}\int r^{2}\mathcal{Q}_{1,-1}^{\left(  0\right)
}\left(  r\right)  J_{0}(k_{z}r)dr\nonumber\\
&  -ik_{\perp}^{1/2}{\sum\limits_{n=1}^{\infty}}\frac{\left(  -1\right)
^{n}6\sqrt{6}}{\mathcal{N}_{2}\left(  n\right)  }\int r^{2}\mathcal{Q}%
_{1,-1}^{\left(  n\right)  }\left(  r\right)  J_{2n}(k_{z}r)dr\nonumber\\
&  -ik_{\perp}^{1/2}{\sum\limits_{n=0}^{\infty}}\frac{\left(  -1\right)
^{n}2\sqrt{6}}{\mathcal{N}_{1}\left(  n\right)  }\int r^{2}j_{1,-1}^{\left(
2\right)  }\left(  r\right)  J_{2n+1}(k_{z}r)dr,
\end{align}%
\begin{align}
H_{\text{int}\left(  \ell=2,\sigma=-1\right)  }^{\text{dip}}  &  =-ik_{\perp
}^{5/2}\frac{\sqrt{6}}{30}\int r^{4}\mathcal{P}_{1,-1}\left(  r\right)
J_{0}(k_{z}r)dr\nonumber\\
&  +ik_{\perp}^{5/2}{\sum\limits_{n=1}^{\infty}}\frac{\left(  -1\right)
^{n}15\sqrt{6}}{\mathcal{N}_{4}\left(  n\right)  }\int r^{4}\mathcal{P}%
_{1,-1}\left(  r\right)  J_{2n}(k_{z}r)dr,
\end{align}%
\begin{align}
H_{\text{int}\left(  \ell=-1,\sigma=1\right)  }^{\text{dip}}  &  =-k_{\perp
}^{3/2}{\sum\limits_{n=0}^{\infty}}\frac{\left(  -1\right)  ^{n}6\sqrt{3}%
}{\mathcal{N}_{3}\left(  n\right)  }\int r^{3}\mathcal{P}_{1,0}\left(
r\right)  J_{2n+1}(k_{z}r)dr\nonumber\\
&  +k_{\perp}^{3/2}\frac{\sqrt{3}}{3}\int r^{3}j_{1,0}^{\left(  2\right)
}\left(  r\right)  J_{0}(k_{z}r)dr\nonumber\\
&  +k_{\perp}^{3/2}{\sum\limits_{n=1}^{\infty}}\frac{\left(  -1\right)
^{n}6\sqrt{3}}{\mathcal{N}_{2}\left(  n\right)  }\int r^{3}j_{1,0}^{\left(
2\right)  }\left(  r\right)  J_{2n}(k_{z}r)dr,
\end{align}%
\begin{align}
H_{\text{int}\left(  \ell=1,\sigma=-1\right)  }^{\text{dip}}  &  =k_{\perp
}^{3/2}{\sum\limits_{n=0}^{\infty}}\frac{\left(  -1\right)  ^{n}6\sqrt{3}%
}{\mathcal{N}_{3}\left(  n\right)  }\int r^{3}\mathcal{P}_{1,0}\left(
r\right)  J_{2n+1}(k_{z}r)dr\nonumber\\
&  +k_{\perp}^{3/2}\frac{\sqrt{3}}{3}\int r^{3}j_{1,0}^{\left(  2\right)
}\left(  r\right)  J_{0}(k_{z}r)dr\nonumber\\
&  +k_{\perp}^{3/2}{\sum\limits_{n=1}^{\infty}}\frac{\left(  -1\right)
^{n}6\sqrt{3}}{\mathcal{N}_{2}\left(  n\right)  }\int r^{3}j_{1,0}^{\left(
2\right)  }\left(  r\right)  J_{2n}(k_{z}r)dr,
\end{align}%
\begin{align}
H_{\text{int}\left(  \ell=-2,\sigma=1\right)  }^{\text{dip}}  &  =ik_{\perp
}^{5/2}\frac{\sqrt{6}}{30}\int r^{4}\mathcal{P}_{1,1}\left(  r\right)
J_{0}(k_{z}r)dr\nonumber\\
&  -ik_{\perp}^{5/2}{\sum\limits_{n=1}^{\infty}}\frac{\left(  -1\right)
^{n}15\sqrt{6}}{\mathcal{N}_{4}\left(  n\right)  }\int r^{4}\mathcal{P}%
_{1,1}\left(  r\right)  J_{2n}(k_{z}r)dr,
\end{align}%
\begin{align}
H_{\text{int}\left(  \ell=0,\sigma=-1\right)  }^{\text{dip}}  &  =ik_{\perp
}^{1/2}\frac{\sqrt{6}}{3}\int r^{2}\mathcal{Q}_{11}^{\left(  0\right)
}\left(  r\right)  J_{0}(k_{z}r)dr\nonumber\\
&  +ik_{\perp}^{1/2}{\sum\limits_{n=1}^{\infty}}\frac{\left(  -1\right)
^{n}6\sqrt{6}}{\mathcal{N}_{2}\left(  n\right)  }\int r^{2}\mathcal{Q}%
_{11}^{\left(  n\right)  }\left(  r\right)  J_{2n}(k_{z}r)dr\nonumber\\
&  -ik_{\perp}^{1/2}{\sum\limits_{n=0}^{\infty}}\frac{\left(  -1\right)
^{n}2\sqrt{6}}{\mathcal{N}_{1}\left(  n\right)  }\int r^{2}j_{1,1}^{\left(
2\right)  }\left(  r\right)  J_{2n+1}(k_{z}r)dr,
\end{align}
where we denoted the combinations of the multipole coefficients as%
\begin{align}
\mathcal{P}_{\ell m}\left(  r\right)   &  =j_{\ell m}^{\left(  r\right)
}\left(  r\right)  -j_{\ell m}^{\left(  1\right)  }\left(  r\right)
,\nonumber\\
\mathcal{Q}_{\ell m}^{\left(  n\right)  }\left(  r\right)   &  =j_{\ell
m}^{\left(  r\right)  }\left(  r\right)  -\frac{2}{3}\left(  2n^{2}-3\right)
j_{\ell m}^{\left(  1\right)  }\left(  r\right)  ,
\end{align}
and%
\begin{align}
\mathcal{N}_{1}\left(  n\right)   &  =\left(  2n-1\right)  \left(
2n+3\right)  ,\nonumber\\
\mathcal{N}_{2}\left(  n\right)   &  =\left(  2n-3\right)  \left(
2n-1\right)  \left(  2n+1\right)  \left(  2n+3\right)  ,\nonumber\\
\mathcal{N}_{3}\left(  n\right)   &  =\left(  2n-3\right)  \left(
2n-1\right)  \left(  2n+3\right)  \left(  2n+5\right)  ,\nonumber\\
\mathcal{N}_{4}\left(  n\right)   &  =\left(  2n-5\right)  \left(
2n-3\right)  \left(  2n-1\right)  \left(  2n+1\right)  \left(  2n+3\right)
\left(  2n+5\right)  .
\end{align}
We summarize the allowed absorptions as follows,%
\begin{equation}
\left(  J,\ell,\sigma\right)  =\left\{
\begin{array}
[c]{c}%
\left(  1,0,1\right) \\
\left(  1,2,-1\right) \\
\left(  0,-1,1\right) \\
\left(  0,1,-1\right) \\
\left(  -1,-2,1\right) \\
\left(  -1,0,-1\right)
\end{array}
\right.  . \label{Selection_rule_VSH}%
\end{equation}
In other words, the absorptions are allowed in case of the optical TAM, $J=1$,
$0$, and $-1$. We note that the selection rule in Eq.
(\ref{Selection_rule_VSH}) includes our result in text, $J=1$. We can say that
this result is consistent with that in text.

\end{document}